\documentstyle[12pt]{article}

\newcommand{\qed}{\hbox{\rule{5pt}{8pt}}}

\catcode `\@=11
\def\@cite#1#2{#1\if@tempswa , #2\fi}
\catcode `\@=12

\newcommand{\upcite}[1]{${}^{\mbox{\scriptsize (\cite{#1})} }$}

\newcommand{\uprcite}[2]{${}^{\mbox{\scriptsize (\cite{#1} - \cite{#2})} }$}
\newcommand{\h}{{\cal H}}
\newcommand{\A}{{\cal A}}
\renewcommand{\L}{{\cal L}}

\begin{document}

\begin{center}
{\Large\bf Quantum fluctuations in quantum lattice-systems}
\smallskip\\
{\Large\bf with continuous symmetry\footnote{
To appear in J.~Stat.~Phys.}} 
\bigskip
\bigskip

{\large Tsutomu Momoi\footnote{E-mail: momoi@cm.ph.tsukuba.ac.jp.}}
\bigskip
\smallskip

{\it Institute of Physics, University of Tsukuba, 
Tsukuba, Ibaraki 305, Japan.\/}
\bigskip
\smallskip

(\hspace{5cm})\\
\end{center}
%\bigskip

\begin{abstract}
\begin{normalsize}
\noindent
We discuss conditions for the absence of spontaneous breakdown of 
continuous symmetries in quantum lattice systems at $T=0$. Our analysis 
is based on Pitaevskii and Stringari's idea that the uncertainty relation 
can be employed to show quantum fluctuations. For the one-dimensional 
systems, it is shown that the ground state is invariant under the continuous 
transformation if a certain uniform susceptibility is finite. 
For the two- and three-dimensional systems, it is shown that 
truncated correlation functions cannot decay 
any more rapidly than $|r|^{-d+1}$ 
whenever the continuous symmetry is spontaneously broken. 
Both of these phenomena occur owing to quantum fluctuations. 
Our theorems cover a wide class of quantum lattice-systems having 
not-too-long-range 
interactions. 
\end{normalsize}
\end{abstract}
\smallskip

uncertainty relation; clustering.
\smallskip

\section{Introduction}
\hspace*{\parindent}
It is well known that continuous symmetries 
cannot be spontaneously broken in one- and two-dimensional systems 
at nonzero temperatures if the interactions are short range. 
Since Mermin and Wagner,\upcite{MerminW} and 
Hohenberg\upcite{Hohenberg} showed rigorous proofs, several 
papers have appeared, proving the invariance of the 
state under the continuous transformation.\uprcite{DobrushinS}{BonatFPK} 
These arguments, however, work only at finite temperatures. 

Absence of symmetry breaking in the ground state of the 
one-dimensional quantum systems has been a long-standing question. This 
problem was discussed by using an extension of the Bogoliubov 
inequality\upcite{Takada} and using the uncertainty 
relation.\upcite{PitaevskiiS,Shastry} Takada\upcite{Takada} 
argued the relation between the absence of long-range order and the 
dispersion form of the excitation 
spectrum, and thereby showed that, 
if the lowest excitation frequency has a gapless 
$k$-linear form, the ground state cannot show symmetry breaking. 
Pitaevskii and Stringari\upcite{PitaevskiiS} 
proposed a zero-temperature analogue of the Bogoliubov inequality, 
using the uncertainty relation
of the quantum mechanics. They presented a method for 
showing the absence of breakdown of continuous symmetry in the ground 
state. After that, 
Shastry\upcite{Shastry} pointed out that one can complete the proof 
for the one-dimensional Heisenberg antiferromagnet 
combining their method and the infrared bound given by 
Dyson, Lieb and Simon.\upcite{DysonLS} 
The method proposed by Pitaevskii and Stringari\upcite{PitaevskiiS} 
can be successfully applied only when we have a rigorous upper bound of 
the susceptibility at the whole momentum space. It is however 
difficult (to date) to obtain upper bounds of the momentum-dependent 
susceptibility in general quantum systems. 

Another well-known theorem for short-range systems 
with continuous symmetry is the 
Nambu-Goldstone theorem, which states that 
there exist gapless elementary excitations whenever any continuous symmetry 
is spontaneously broken. 
This theorem was also proved for the lattice systems.\upcite{LandauFPW} 
Furthermore, Martin\upcite{Martin} showed that some truncated correlation 
functions at finite temperatures have power-decay behavior 
slower than or equal to $|r|^{-1}$ 
in three-dimensional systems if continuous symmetry is spontaneously 
broken. 

In the present paper, we extend the method by Pitaevskii and 
Stringari,\upcite{PitaevskiiS} using the technique developed 
by Martin,\upcite{Martin} 
and thereby show conditions for the ground state of quantum systems 
being invariant under the continuous transformation. 
We obtain the following results on the continuous-symmetry breaking 
in the ground states. 
\begin{itemize}
\item[1.] In the one-dimensional system, if a certain uniform 
susceptibility is finite, the ground state has continuous symmetry, i.e.,
\begin{equation}
\omega ( \sigma_\theta (A)) = \omega ( A )
\end{equation}
for any local observable $A$. Here $\omega ( \cdots )$ denotes the 
ground state and $\sigma_\theta$ denotes 
the continuous transformation under which 
the interactions of the Hamiltonian are invariant. (See theorem 1.)
\item[2.] In more-than-one-dimensional ($d>1$) systems, if any continuous 
symmetry is spontaneously broken in the ground state as 
${d\over d\theta} \omega ( \sigma_\theta (A) ) |_{\theta=0} \ne 0$ 
with a local observable $A$ 
and if a certain uniform susceptibility is finite, the truncated two-point 
correlation function 
of $A$ shows a power-decay slower than or equal 
to $O(1/r^{d-1})$. Here we denote the 
dimensionality of the system by $d$. (See theorem 2.)
\end{itemize}
Both of these phenomena occur as a consequence of quantum fluctuations. 
In our discussion, we define the ground state applying an infinitesimally 
small field. 
We derive these results, using rigorous inequalities and  
assuming the clustering property of this ground state. 
(Note that this assumption is quite reasonable, though it cannot be verified 
within the presently available techniques in mathematical physics.)
These theorems are applicable to a wide class of quantum lattice systems 
having 
not-too-long-range interactions and continuous symmetries. Quantum spin 
systems, lattice fermion-systems and hard-core bose systems 
are included, for example. 

\section{Theorems and physical consequences} 
\subsection{Preliminaries}
We first give some notations. We denote the $d$-dimensional lattice by 
$\L$, which 
is taken as ${\bf Z}^d$. For each lattice point $x\in\L$, 
there are the algebra 
$\A_x$ of operators and the finite-dimensional Hilbert space $\h_x$. For any 
bounded subset $\Lambda\subset\L$, local operators which are defined 
on $\Lambda$ generate the local algebra $\A_\Lambda$ of 
observables and the Hilbert space is given 
by $\h_\Lambda = \bigotimes_{x\in\Lambda} \h_x$. 

For simplicity, we present arguments for quantum systems with two-body 
interactions. We can easily extend the following arguments to models with 
more-than-two-body interactions. 
Let $\L$ be the translationally invariant lattice and 
$H_\Lambda$ be the Hamiltonian in the finite-volume lattice 
$\Lambda\subset\L$, which is given by 
\begin{equation}
H_\Lambda = \sum_{x,y\in \Lambda} \phi(x,y).
\end{equation}
Here $\phi(x,y)$ denotes the translationally invariant interaction 
defined on ${\h}_x \otimes {\h}_y$ 
with the norm $\Vert \phi(x,y)\Vert = \psi (x-y)$. We restrict our discussions 
to the models that have not-too-long-range interactions satisfying 
\begin{equation}\label{eq:range}
\sum_{x\in\L} |x|^2 \psi (x) <\infty 
\end{equation}
and that have, at least, 
the  $U(1)$-continuous symmetry, i.e., 
\begin{equation}\label{eq:symmetry}
[\phi (x,y), J_\Omega ]=0
\end{equation}
for any $x$, $y\in \Omega$ and local subset $\Omega\subset\L$. 
Here $J_\Omega$ denotes the generator 
of the (global) symmetry-transformations of operators in $\A_\Omega$. 
The continuous symmetry-transformation is given by 
\begin{equation}
\sigma_\theta (A) = \exp (i\theta J_\Omega)A\exp (-i\theta J_\Omega)
\end{equation}
for any $A\in \A_\Omega$.

To define the ground state, we select 
a proper order parameter and then apply the corresponding symmetry-breaking 
field. 
Let us define the ground state in the form 
\begin{equation}\label{eq:GS}
\omega ( \cdots ) 
 = \lim_{B\downarrow 0} \lim_{\Lambda \uparrow \L} 
   \lim_{\beta\uparrow\infty}
     {{\rm Tr} \cdots \exp \{-\beta (H_\Lambda -BO_\Lambda) \} 
       \over  {\rm Tr}\exp \{-\beta (H_\Lambda -BO_\Lambda) \}}, 
\end{equation}
where $O_\Lambda$ denotes the order-parameter operator and $B$ is the 
real-valued symmetry breaking field. It is known that the limits are 
well-defined by choosing suitable sequences of $\Lambda$ and $B$. 
(See Appendix A of ref.~\cite{KomaT1994}, for example.) 

We restrict our discussions to the case that the order-parameter operator has 
a sublattice-translational invariance. Hence the ground state defined by 
(\ref{eq:GS}) has the following sublattice-translational invariance 
\begin{equation}\label{cond:translation}
\omega (A) = \omega (\tau_x (A))
\end{equation}
for any $x\in \L_{\rm s}$ and $A\in \A_\Omega$ on a local subset $\Omega$. 
Here $\tau_x$ denotes the space translation by $x$ and 
$\L_{\rm s}$ denotes a set of sites in a sublattice. 
If we consider antiferromagnets on a bipartite lattice, for example, 
the order parameter 
is set as the staggered magnetization and $\L_{\rm s}$ is one of two 
sublattices. In ordinary ferromagnets, the ground state 
has the full lattice-translational invariance and hence $\L_{\rm s}=\L$. 

In the following discussions, we assume the clustering property of the state, 
\begin{equation}\label{cond:cluster}
|\omega ( \tau_x (A) B ) 
  - \omega ( \tau_x (A) ) \omega ( B ) | 
\le O \left( {1 \over |x|^\delta} \right)
\end{equation}
with $\delta>0$ for sufficiently large $|x|$ and any $A$, $B\in\A_\Omega$ 
on a local subset 
$\Omega$. 
This property means that observations at two points separating far away from 
one another do not affect each other. Note that this is a quite 
natural assumption. It is believed that, 
by selecting a proper order parameter, the state $\omega ( \cdots )$ 
becomes an pure state, i.e., it has the clustering property. 

{\bf Remark:} It is widely believed that any physically 
natural equilibrium state has the clustering property. 
In studies on finite systems, we sometimes encounter states 
which do not have the cluster property. For example, consider the 
ground state of the three-dimensional Heisenberg antiferromagnet. 
It is shown that the ground state 
of finite-volume systems is invariant under the global spin 
rotation\upcite{Marshall,LiebM} and it has 
a long-range order in the infinite-volume limit.\upcite{KennedyLS} 
Taking the infinite-volume limit of the ground state of finite systems, one 
can define a ground state that does not have the clustering property. 
However, as discussed in ref.~\cite{KomaT1994}, this 
symmetric ground state is unphysical and only a mathematical object. 
It is believed that in the thermodynamic limit 
this state is decomposed into pure states 
and one of the pure states appears as a natural state in the real 
system.\upcite{Ruelle,BratteliR} 
%\newpage

\subsection{Main Theorems}
In this section, we show our theorems. Physical 
consequences of the theorems will be discussed in sections~\ref{sec:one-d} 
and \ref{sec:two-three-d}, and proofs are given in section~\ref{sec:proof}. 

The statement that the state $\omega ( \cdots )$ 
has the continuous symmetry is equivalent to 
\begin{equation}
{d\over d\theta} \omega ( \sigma_\theta (A) ) \biggr|_{\theta =0} = 0
\end{equation}
for any $A\in \A_\Lambda$ on any subset $\Lambda \subset \L$. 
We consider the transformation $\sigma_\theta$ in which $J_\Omega$ is 
given by $J_\Omega = \sum_{x\in \Omega}\tau_x (J_0)$ with a bounded 
self-adjoint operator $J_0\in\A_0$. 
In this case, we have 
\begin{equation}\label{eq:deriv}
{d\over d\theta} \omega ( \sigma_\theta (A) ) \biggr|_{\theta =0}
 = i\omega ( [J_\Lambda,A] ) 
\end{equation}
for any $A\in\A_\Lambda$ and any subset $\Lambda\subset\L$. 

Without loss of generality, we consider the operator $A$ on the subset 
$\Lambda = \{\mbox{$x\in\L:$ $|x|\le r$}\}$, where $r$ is a finite constant. 
To discuss the quantity $\omega ( [J_\Lambda,A] )$, 
we use the sublattice-translational invariance (\ref{cond:translation}) 
and hence we have 
\begin{equation}\label{eq:translation}
\omega ( [J_\Lambda,A] ) 
 = {1\over |\Omega_{\rm S}|}\sum_{x\in \Omega_{\rm S}} 
    \omega ( [J_\Omega,\tau_x (A)] )
\end{equation}
for any $A\in\A_\Lambda$, 
where $\Omega_{\rm S} = \{\mbox{$x\in\L_{\rm S}:$ $|x|\le R$} \}$ and 
$\Omega = \{\mbox{$x\in\L:$ $|x_i|\le R_0$ for $i=1,\dots ,d$} \}$ with 
$R_0 = R+r$. [Though equation (\ref{eq:translation}) holds by 
setting $\Omega$ as $\{ x\in \L$~:~$|x|\le R+r \}$, we have taken $\Omega$ 
as the hyper-cubic lattice for convenience in later discussions.] 
Bounding the absolute value of the right-hand side of~(\ref{eq:translation}) 
with the uncertainty relation and the Kennedy-Lieb-Shastry 
inequality\upcite{KennedyLS}, and 
estimating the $R$ dependence of the upper bound, 
we obtain the following lemma. 
%\vspace{12mm}
\vspace{6mm}

{\bf Lemma.} Let the interaction satisfy~(\ref{eq:range}) 
and (\ref{eq:symmetry}), and assume that the ground state (\ref{eq:GS}) 
satisfies (\ref{cond:translation}) and (\ref{cond:cluster}). 
Consider $\A_\Lambda$ on the subset $\Lambda=\{ x\in \L : |x| \le r \}$, 
where $r$ is a finite constant, and let 
$\Omega_{\rm S} = \{\mbox{$x\in\L_{\rm S}:$ $|x|\le R$} \}$ and 
$\Omega = \{\mbox{$x\in\L:$ $|x_i|\le R_0$ for $i=1,\dots,d$} \}$ with 
$R_0=R+r$. 
Furthermore, define the uniform susceptibility of $J$ by 
\begin{equation}\label{def:suscep2}
\chi_J = \lim_{\Omega \uparrow \L} {2 \over |\Omega|} 
         \int^\infty_0 d\lambda 
        \{ \omega( J_\Omega J_\Omega(i\lambda) ) - \omega^2( J_\Omega ) \}
       \ge 0
\end{equation}
assuming existence of the limit,\footnote{
Mathematically speaking, existence of the limit in (\ref{def:suscep2}) 
may be nontrivial. It should 
be remarked that this definition of the uniform susceptibility is 
equivalent to the standard one, which has been used in many papers 
in physics. See Appendix.} 
where $J_\Omega (t)$ is the time-evolved operator of $J_\Omega$. 
Then, the right-hand side of (\ref{eq:translation}) is bounded as 
\begin{equation}\label{Lemma}
\left|{1 \over |\Omega_{\rm s}|} \sum_{x\in \Omega_{\rm s}} 
       \omega ([J_\Omega, \tau_x (A)] )  \right|^2 
 \le \left\{ 
\begin{array}{ll}
O(R^{d-1-\delta})\cdot\sqrt{\chi_{J}} & 
\mbox{\hspace{1cm}} (0<\delta<d) \\%[0.5cm]
O(R^{-1}\ln R)\cdot\sqrt{\chi_{J}}    & 
\mbox{\hspace{1cm}} (\delta=d)   \\%[0.5cm]
O(R^{-1})\cdot\sqrt{\chi_{J}}         & 
\mbox{\hspace{1cm}} (\delta>d) 
\end{array} \right.
\end{equation}
for sufficiently large $R$ and any $A\in\A_\Lambda$. 
%\vspace{12mm}
\vspace{6mm}

We will give a proof in section~\ref{sec:proof}. As shown in the proof, 
this lemma comes from the uncertainty relation of the quantum mechanics. 
Hence the inequality (\ref{Lemma}) can show purely quantum effects. 
In the following, 
we discuss physical consequences of the bound in each dimension. 
It should be remarked that these results are applicable to various 
models on arbitrary lattices that have the translation invariance. 
Selecting bonds of the non-vanishing interactions $\phi (x,y)$, 
we can define various lattices on ${\bf Z}^d$. 
The results depend only on the dimensionality $d$ of the lattice. 

First, we discuss one-dimensional systems, in which $\L={\bf Z}$. 
By taking the $R\rightarrow\infty$ limit of (\ref{Lemma}), 
the above lemma shows conditions for the absence 
of continuous-symmetry breaking in one-dimensional systems. 
%\vspace{12mm}
\vspace{6mm}

{\bf Theorem 1.} Let $\L$ be a one-dimensional lattice and the 
interaction $\phi (x,y)$ 
satisfy~(\ref{eq:range}) and (\ref{eq:symmetry}). Assume the ground 
state (\ref{eq:GS}) satisfies the properties (\ref{cond:translation}) 
and (\ref{cond:cluster}). If the infinite volume limit in the 
definition (\ref{def:suscep2}) of the uniform susceptibility exists and if 
this 
susceptibility $\chi_J$ is not diverging, the ground state (\ref{eq:GS}) 
is invariant under the continuous transformation $\sigma_\theta$, i.e.,
\begin{equation}\label{eq:Theorem1}
{d \over d\theta}\omega ( \sigma_\theta (A) ) \biggr|_{\theta=0} = 0
\end{equation}
for any $A\in\A_\Lambda$ on any finite subset $\Lambda$. 
%\vspace{12mm}
\vspace{6mm}

Physical meanings of this theorem are discussed in 
section~\ref{sec:one-d}. 
An advantageous point of this theorem is that the results depend only on the 
``uniform'' susceptibility, not on other momentum-dependent 
susceptibilities. 
The condition that the uniform susceptibility is 
finite (or vanishing) is physically important. (See examples in the 
next section.) We cannot improve the condition 
without further detailed properties of the model, 
since the uniform susceptibility is finite or 
diverging, depending on each model. 

Next, we discuss two- and three-dimensional systems. 
For these systems, we consider the case that the continuous symmetry is 
spontaneously broken. Slight modifications of the lemma 
give the following bound for a truncated two-point correlation function. 
%\vspace{12mm}
\vspace{6mm}

{\bf Theorem 2.} Let $\L$ be a more-than-one-dimensional ($d>1$) lattice, 
and $\phi (x,y)$ satisfy~(\ref{eq:range}) and (\ref{eq:symmetry}). 
Assume that the ground state (\ref{eq:GS}) satisfies the conditions 
(\ref{cond:translation}) and (\ref{cond:cluster}). 
If continuous symmetry is spontaneously broken in the ground 
state (\ref{eq:GS}), i.e.,  $\omega ( [J_\Lambda,A] ) \ne 0$ 
with an operator $A\in \A_\Lambda$ on an arbitrary subset 
$\Lambda \subset \L$, and  if the infinite volume limit in 
(\ref{def:suscep2}) exists and $\chi_J <\infty$, 
the truncated two-point correlation function of $A$ shows the slow 
clustering as 
\begin{equation}\label{eq:Theorem2}
|\omega ( A^* \tau_x (A) ) 
  - \omega ( A^* ) \omega ( \tau_x (A) ) | 
\ge O\left( {1\over |x|^{d-1}} \right)
\end{equation}
for sufficiently large $|x|$ with $x\in\L_{\rm S}$. 
%\vspace{12mm}
\vspace{6mm}

We discuss the meaning of this theorem in section~\ref{sec:two-three-d} 
and give a proof in section~\ref{sec:proof}. 
Under some conditions, this theorem states that the truncated
correlation function of $A$ cannot show 
any exponential decay in the ordered ground state. 
This result hence corresponds to an extension 
of the Nambu-Goldstone theorem. 
This theorem shows the conditions for existence of quantum fluctuations 
and shows strong correlation between the fluctuations. 
(Remember that in the classical model there is no fluctuation in the 
ground state and hence the truncated two-point correlation function 
vanishes.)
The condition for the uniform susceptibility appears in 
the theorem again and it is important in this case, as well. (See 
examples in section~\ref{sec:two-three-d}.)
%\newpage

\subsection{One-dimensional systems}
\label{sec:one-d}
First we discuss one-dimensional systems, whose lattice is set as 
${\bf Z}$. Among the assumptions of Theorem~1, the finiteness of $\chi_J$ 
is physically important. It determines whether the ground state shows 
symmetry breaking or not. 
To clarify the meaning of Theorem~1, we display three examples. 

{\bf Example 1.} {\em Spin SU(2) symmetry.} 
Let us first consider the one-dimensional spin-$S$ Heisenberg antiferromagnet 
on the lattice $\L(={\bf Z})$. The Hamiltonian in $\Lambda\subset\L$ 
is given by 
\begin{equation}\label{eq:Heisenberg}
H_\Lambda = \sum_{\langle i,j \rangle \in \Lambda} 
(S_i^x S_j^x + S_i^y S_j^y + S_i^z S_j^z),
\end{equation}
where $S^\alpha_i$ ($\alpha=x$, $y$, $z$) denote the spin operators on the 
site $i$ satisfying 
$[S^\alpha_j,S^\beta_k]=i\delta_{jk}\epsilon_{\alpha\beta\gamma}S^\gamma_j$
with ${\bf S}^2=S(S+1)$. 
The summation runs over all the nearest-neighbor sites. 
As a generator of the $U(1)$ rotation, we take 
\begin{equation} 
J_\Lambda = \sum_{i\in\Lambda} S^z_i. 
\end{equation} 
This model clearly satisfies the conditions~(\ref{eq:range}) 
and (\ref{eq:symmetry}). 
Setting the order-parameter operator of the antiferromagnetism as 
\begin{equation}
O_\Lambda = \sum_{i\in\Lambda} (-1)^i S^x_i, 
\end{equation}
we define the ground state $\omega(\cdots)$ by (\ref{eq:GS}). 
By definition, the ground state satisfies the sublattice-translation 
invariance (\ref{cond:translation}). 
In this model, the quantity $\chi_J$ is the uniform magnetic 
susceptibility of the ground state $\omega(\cdots)$. 
It has been proved 
in refs.~\cite{DysonLS} and \cite{KennedyLS} that $\chi_J$ is bounded from 
above by a finite constant for the Heisenberg antiferromagnets 
on hyper-cubic lattices.\footnote{\label{footnote}
Though, in refs.~\cite{DysonLS} and \cite{KennedyLS}, they discussed only 
antiferromagnets without any magnetic field, their arguments can be easily 
extended to the Hamiltonian with the staggered magnetic field, 
$H_\Lambda - B O_\Lambda$, and hence we can show that their bound on the 
susceptibility holds for this system as well.} 
Finally we assume that $\omega(\cdots)$ satisfies the clustering property. 
Under this assumption, Theorem~1 hence states that 
the ground state $\omega(\cdots)$ has the spin-rotational symmetry. 

For the system whose uniform susceptibility is not diverging, 
Theorem 1 states that 
quantum fluctuations suppress spin ordering, even if some 
momentum-dependent susceptibility is diverging. 
The correlation of $k=0$ is however special. Theorem~1 does not 
exclude the possibility of ferromagnetism, since in the ferromagnets 
the uniform transverse susceptibility, which is nothing but $\chi_J$, 
diverges. As is well known, 
the one-dimensional Heisenberg ferromagnet has the fully ordered ground 
state. Thus the spin long-range correlation with the zero momentum can survive 
quantum fluctuations. Furthermore Theorem 1 says that 
ferrimagnetism can occur as well. 
Some models indeed show the ferrimagnetic order even 
in the one-dimensional system.\upcite{ShenQ} 
In the ferrimagnetism, antiferromagnetic long-range 
order coexists with ferromagnetic order. From 
the theorem we learn that this antiferromagnetic order 
can appear owing to the existence of ferromagnetic order. 

{\bf Example 2.} {\em Spin O(2) symmetry.} 
Next we consider the one-dimensional spin-$S$ $XY$ ferromagnet, 
whose Hamiltonian is 
\begin{equation}
H=-\sum_{\langle i,j \rangle\in\Lambda} (S^x_i S^x_j + S^y_i S^y_j).
\end{equation}
The summation runs over all nearest-neighbor sites. This model is 
invariant under the $O(2)$ rotation, whose generator is 
$J_\Lambda=\sum_{i\in\Lambda} S^z_i$. 
This model is expected to have strong correlation at $k=0$. 
We hence set the order parameter as $O_{\Lambda}=\sum_{i\in\Lambda}S^x_i$, 
thereby defining the ground state by (\ref{eq:GS}). 
The Hamiltonian and the ground state clearly satisfy the conditions 
(\ref{eq:range}), (\ref{eq:symmetry}) and (\ref{cond:translation}).
Since this model has only the $O(2)$ symmetry and may have 
weak $S^z$-correlation, situations are different from the 
ferromagnets in the above example. 
We expect $\chi_J$ is not diverging in the $XY$ model and hence, from 
Theorem~1, the ground state has the $O(2)$ rotational invariance. 

{\bf Example 3.} {\em U(1)-gauge symmetry of fermions.} 
Let us consider the breakdown of the $U(1)$-gauge 
symmetry of fermions. (The Hilbert space of fermion systems is not 
a simple tensor product of the local Hilbert spaces and hence 
some modifications 
to the notations are needed. Furthermore, each observable in the algebra 
$\A_\Omega$ should contain multiplets of an even 
number of fermion operators, so that $[A,B]=0$ for 
any $A\in\A_{\Lambda_1}$ and 
$B\in\A_{\Lambda_2}$ with $\Lambda_1 \cap \Lambda_2 =\emptyset$. Thereby 
our theorems still work for the fermion systems, 
as well.) 

As an example of correlated lattice-fermions, 
we consider the one-dimensional Hubbard model, whose Hamiltonian is given by 
\begin{equation}\label{eq:Hubbard}
H_\Lambda 
 = - t \sum_{\langle i,j \rangle \in \Lambda} 
       \sum_{\sigma=\uparrow,\downarrow} 
        (c_{i\sigma}^* c_{j\sigma} + c_{j\sigma}^* c_{i\sigma}) 
   + U\sum_{i\in\Lambda} n_{i\uparrow}n_{i\downarrow}
   -\mu \sum_{i\in\Lambda} (n_{i\uparrow} + n_{i\downarrow}). 
\end{equation}
The summation of the hopping term runs over all the neatest-neighbor sites. 
We denote the creation operator of the fermion at site $i$ 
with spin $\sigma$ by 
$c_{i\sigma}^*$ and the number operator of the fermion by $n_{i\sigma}$. 
The generator of the gauge transformation is given by 
$J_\Lambda = \sum_{i\in\Lambda} (n_{i\uparrow}+n_{i\downarrow})$ 
and hence $\chi_J$ is the uniform charge 
susceptibility, or the compressibility. This model satisfies the conditions 
(\ref{eq:range}) and (\ref{eq:symmetry}). Under the assumption of the 
clustering property, Theorem~1 states for this model that, if the 
compressibility is finite, there is no breakdown of the 
$U(1)$-gauge symmetry. 

Here we mention about the model proposed by Essler et 
al.\upcite{EsslerKS} In their model the 
ground state has superconductivity even in the one-dimensional 
system. It should be noted that the compressibility is 
diverging in the ground state of their model, 
and hence Theorem~1 is not applicable to their model. 
%\newpage

\subsection{Two- and three-dimensional systems} 
\label{sec:two-three-d}
In this section we discuss two- and three-dimensional systems, whose 
lattice is taken as ${\bf Z}^2$ or ${\bf Z}^3$. 
To clarify the meaning of Theorem~2, let us consider two examples. 

{\bf Example 4.} We again discuss the spin-symmetry breaking of the Heisenberg 
antiferromagnet (\ref{eq:Heisenberg}). 
Here we take the lattice $\L$ as ${\bf Z}^2$ or 
${\bf Z}^3$. We set the order-parameter operator as 
$O_\Lambda = \sum_{r\in\Lambda} S^x_r 
             \exp (i{\bf q}\cdot {\bf r})$ 
with ${\bf q}=(\pi,\ldots,\pi)$ and the generator of rotation as 
$J_\Lambda = \sum_{r\in\Lambda} S^z_r$. 
This model hence satisfies the conditions (\ref{eq:range}) 
and (\ref{eq:symmetry}), and the ground state $\omega (\cdots)$ defined by 
(\ref{eq:GS}) satisfies (\ref{cond:translation}). 
The occurrence of symmetry breaking in $\omega (\cdots)$
is proved for the two-dimensional $S\ge 1$ 
models\upcite{NevesP,KomaT1993}
and for the three-dimensional arbitrary-$S$ 
models.\upcite{KennedyLS,KomaT1993} 
Existence of long-range order is also proved for anisotropic Heisenberg 
antiferromagnets.\uprcite{KennedyLS2}{NishimoriO} 
In these models, the ground state hence shows 
\begin{equation}
\omega ( [J_\Lambda , S^y_r] ) = -i \omega ( S^x_r ) 
\ne 0.
\end{equation}
Furthermore, the finiteness of $\chi_J$ is proved in 
refs.~\cite{DysonLS} and \cite{KennedyLS}. 
(See also the footnote~3 on page~\pageref{footnote}.) 
Using these results and assuming the clustering property of 
$\omega(\cdots)$, we find from Theorem~2 that 
the transverse-spin correlation shows the slow clustering as 
\begin{equation}\label{eq:spin_correlation}
|\omega ( S^y_0 S^y_r )| \ge O\left( {1\over |r|^{d-1}} \right)
\end{equation}
for $0$,~$r\in\L_{\rm S}$ and for sufficiently large $|r|$. 
Note that $\omega(S^y_r)=0$ by definition of the ground state. 
Hence (\ref{eq:spin_correlation}) shows 
that there are quantum fluctuations in the ground state 
and they are strongly correlated. 
Shastry\upcite{Shastry} showed that the transverse-structure factor 
diverges as $\omega(S^y_k S^y_{-k})\sim 1/|{\bf k}-{\bf q}|$ at 
${\bf k}\simeq {\bf q}$ in the ground state with N\'eel order. 
This indicates that the transverse-correlation function decays as 
$\omega(S^y_0 S^y_r)\sim (-1)^r /|r|^{d-1}$. Thus this example shows that 
our bound (\ref{eq:Theorem2}) is optimal. 

It may be worth mentioning about another Nambu-Goldstone-type theorem for 
the excitation spectrum of the Heisenberg 
antiferromagnets.\uprcite{Momoi1994}{Momoi1995} 
It states that the 
N\'eel-ordered ground state has a gapless excitation spectrum and the 
lowest frequency of excitations is bounded from above by a gapless $k$-linear 
form around ${\bf k}\simeq {\bf 0}$ and {\bf q}. 
These two Nambu-Goldstone-type theorems may closely relate to 
each other. 

Furthermore we discuss 
the ferromagnetic Heisenberg model, in which $\chi_J$ is diverging, 
to clarify the significance of the condition on $\chi_J$. 
The ground state of the ferromagnet can be written as a direct product 
of local spins and it does not fluctuate. Hence the truncated two-point
correlation function is always vanishing. 
Thus the Heisenberg ferromagnet 
is a special model, which does not contain quantum fluctuations in the 
ground state. Our theorem successfully excludes this special case. 

{\bf Example 5.} Finally we consider lattice fermion-systems, e.g., 
the Hubbard model (\ref{eq:Hubbard}) on ${\bf Z}^2$ or ${\bf Z}^3$, 
and consider the spontaneous breakdown of the $U(1)$-gauge symmetry. 
As the order parameter, we take, for example, 
\begin{equation}
O_\Lambda = \sum_{i\in\Lambda}O^+_i 
 = \sum_{i\in\Lambda} (c_{i\uparrow}^* c_{i\downarrow}^* 
                       + c_{i\downarrow} c_{i\uparrow}).
\end{equation}
One can take other types of order parameters, as well. 
The generator of gauge transformation is given by 
$J_\Lambda=\sum_{i\in\Lambda} (n_{i\uparrow} + n_{i\downarrow})$ and hence 
$\chi_J$ denotes the charge susceptibility. Assume that the ground state 
defined by (\ref{eq:GS}) shows superconductivity and satisfies 
\begin{equation}
\omega ( [J_\Lambda , O^-_j] ) = 2 i \omega ( O^+_j ) \ne 0, 
\end{equation}
where $O^-_j = i c_{j\uparrow}^* c_{j\downarrow}^* 
- i c_{j\downarrow} c_{j\uparrow}$. 
For this system, Theorem~2 states that, if the compressibility is 
finite, we have 
\begin{equation}
|\omega ( O^-_0 O^-_r )| \ge O\left( {1\over |r|^{d-1}} \right) 
\end{equation}
for sufficiently large $|r|$. 

It should be remarked that the Coulomb interaction does not satisfy 
the condition (\ref{eq:range}) 
and hence Theorem~2 is not applicable to the 
models that contain the Coulomb interactions. Decay of correlation 
functions in these systems may closely relate to the 
Anderson-Higgs phenomena and it is out of scope of this paper. 
%\newpage

\subsection{Proof of Theorems}
\label{sec:proof}

In this section, we shall show proofs of Lemma, Theorem~1 and Theorem~2. 
%\vspace{12mm}
\vspace{6mm}

{\it Proof of Lemma.} 
As in ref.~\cite{PitaevskiiS}, we use the following two 
inequalities; one is the uncertainty relation,\upcite{PitaevskiiS} 
\begin{equation}\label{ineq:Uncertainty}
|\omega ( [C,A])|^2
 \le \omega ( \{ \Delta C^*,\Delta C \} ) 
     \omega ( \{ \Delta A^*,\Delta A \} )
\end{equation}
for any $A$, $C\in\A_\Omega$ with $\Delta C = C - \omega ( C )$ 
and $\Delta A = A - \omega ( A )$, and 
the other is Kennedy, Lieb and Shastry's 
inequality,\upcite{KennedyLS} 
\begin{equation}\label{ineq:KLS}
\omega ( \{\Delta C^*, \Delta C\})^2
 \le D(C) \omega ( [ [ C^*, H_\Omega ], C ] )
\end{equation}
for any $C\in\A_\Omega$. 
Here $H_\Omega$ denotes the Hamiltonian on $\Omega$ and $D(C)$ denotes the 
Duhamel two-point function of $C$, 
\begin{equation}
D(C) = \lim_{B\downarrow 0}\lim_{\Lambda\uparrow\L}
       \lim_{\beta\uparrow\infty}
       \int^\beta_0 d\lambda 
     \{ \omega_{\Lambda,B} ( C^* C( i\lambda ) )
       - \omega_{\Lambda,B} ( C^* ) \omega_{\Lambda,B} ( C ) \} , 
\end{equation}
where 
\begin{equation}
\omega_{\Lambda,B} ( \cdots ) 
 = \frac{ {\rm Tr} [\cdots \exp \{-\beta(H_\Lambda -BO_\Lambda)\}] }
        { {\rm Tr} [\exp \{-\beta(H_\Lambda -BO_\Lambda)\}] } 
\end{equation}
and 
\begin{equation}\label{eq:time_evolution}
C(t) = \exp \{ it (H_\Lambda -BO_\Lambda)\}
        C \exp \{ -it (H_\Lambda -BO_\Lambda)\}. 
\end{equation}
Both inequalities (\ref{ineq:Uncertainty}) and (\ref{ineq:KLS}) 
were first obtained for finite-volume systems. 
Taking the thermodynamic limit of the inequalities, one obtains 
(\ref{ineq:Uncertainty}) and (\ref{ineq:KLS}). Combining 
(\ref{ineq:Uncertainty}) and (\ref{ineq:KLS}), we have 
\begin{equation}
\label{ineq:PitaevskiiS}
|\omega ( [C,A])|^2 
 \le \biggl\{ D(C) \omega ( [ [ C^*, H_\Omega ], C ] )\biggr\}^{1/2}
     \omega ( \{ \Delta A^*, \Delta A\} ) 
\end{equation}
for any $A$, $C\in \A_\Omega$, where $\Delta A = A - \omega ( A )$. 
Setting $A$ as $A_{\Omega_{\rm S}} = |\Omega_{\rm S}|^{-1} 
\sum_{x\in\Omega_{\rm S}} \tau_x (A)$ with $A\in\A_{\Lambda}$ and 
$C=J_\Omega$ in (\ref{ineq:PitaevskiiS}), we obtain an upper 
bound of~(\ref{eq:translation}). 

To estimate properly the $R$ dependence of the right-hand side of 
(\ref{ineq:PitaevskiiS}), we use the smooth 
action\upcite{FrohlichP,KleinLS} of $J_{\Omega}$. 
We set the operator $C$ as 
\begin{equation}
C = J_f = \sum_{x\in\L} f(x) J_x, 
\end{equation}
where $f(x)=1$ for $x\in\Omega$, and $f(x)\rightarrow 0$ as 
$|x|\rightarrow\infty$. 
Defining $x_{\rm max}$ by $x_{\rm max}=\max_i |x_i|$, 
we set the function $f(x)$ in the form 
\begin{equation}
f(x) = \left\{ 
\begin{array}{ll}
1               & \mbox{\hspace{1cm}} (x_{\rm max}<R_0) \\[0.5cm]
2-x_{\rm max}/R_0 
               & \mbox{\hspace{1cm}} (R_0 \le x_{\rm max} \le 2R_0) \\[0.5cm]
0               & \mbox{\hspace{1cm}} (2R_0 <  x_{\rm max}). 
\end{array} \right.
\end{equation}
Hence the operator $C(=J_f)$ is defined on the subset 
$\Omega^\prime = \{x\in\L:$~$|x_i|\le 2R_0$~for~$i=1,\dots,d \}$. 
Thus, we have 
\begin{eqnarray}\label{ineq:1}
|\omega ( [J_{\Omega},A_{\Omega_{\rm S}}])|^2 
 &=& |\omega ( [J_f,A_{\Omega_{\rm S}}])|^2 \nonumber\\
 &\le& \biggl\{ D(J_f) 
                \omega ( [ [ J_f, H_{\Omega^\prime} ], J_f ] ) \biggr\}^{1/2}
       \omega ( \{\Delta A_{\Omega_{\rm S}}^*,\Delta A_{\Omega_{\rm S}}\} ) 
\end{eqnarray}
for any $A\in\A_\Lambda$, where $\Delta A_{\Omega_{\rm S}} 
= |\Omega_{\rm S}|^{-1} \sum_{x\in\Omega_{\rm S}} \tau_x (A) - \omega ( A )$. 
{}From now, we discuss the right-hand side of~(\ref{ineq:1})
estimating the $R$ dependence in the large $R$ limit. 

Let us first discuss $D(J_f)$. The operator $J_f$ can be decomposed as 
\begin{equation}
J_f = {1\over R_0} \sum_{n=0}^{R_0-1} \biggl(\sum_{x\in\Omega(n)} J_x \biggr) 
= {1\over R_0} \sum_{n=0}^{R_0-1} J_{\Omega(n)} ,
\end{equation}
where $\Omega (n)$ denotes the hyper-cubic lattice defined by 
\begin{equation}
\Omega(n) = \{\mbox{$x\in\L:$ $|x_i|\le R_0+n$ for $i=1,\dots,d$}\}. 
\end{equation}
Now we consider the finite-volume lattice $\Lambda (\supset \Omega (R_0))$
and introduce the function 
\begin{equation}
D_{\Lambda,B}(A,C) 
 = \lim_{\beta\uparrow\infty}
   \int^\beta_0 d\lambda \{ 
     \omega_{\Lambda,B} ( A^* C(i\lambda) ) 
     - \omega_{\Lambda,B} ( A^* ) \omega_{\Lambda,B} ( C ) \} 
\end{equation}
for $A,C\in\A_\Lambda$, 
where $C(t)$ is the time-evolved operator of $C$, given in 
(\ref{eq:time_evolution}). 
This function clearly satisfies $D_{\Lambda,B} (A,A) \ge 0$ and the 
linearity $D_{\Lambda,B} (A, a C_1 + b C_2) = 
a D_{\Lambda,B} (A,C_1) + b D_{\Lambda,B} (A,C_2)$ for any $a,b\in{\bf C}$ 
and $A,C_1,C_2\in\A_\Lambda$. We hence regard 
$D_{\Lambda,B} (A,C)$ as the inner product. Inserting $J_f$ into 
$D_{\Lambda,B}$, we obtain 
\begin{eqnarray}
D_{\Lambda,B} (J_f,J_f) 
 &=& {1\over {R_0}^2} \sum_{n=0}^{R_0-1} \sum_{m=0}^{R_0-1} 
        D_{\Lambda,B} (J_{\Omega(n)},J_{\Omega(m)}) \nonumber\\
 &\le& {1\over {R_0}^2} \sum_{n=0}^{R_0-1} \sum_{m=0}^{R_0-1} 
        |D_{\Lambda,B} (J_{\Omega(n)},J_{\Omega(m)})| \nonumber\\
 &\le& {1\over {R_0}^2} \sum_{n=0}^{R_0-1} \sum_{m=0}^{R_0-1} 
  \{ D_{\Lambda,B} (J_{\Omega(n)},J_{\Omega(n)}) 
     D_{\Lambda,B} (J_{\Omega(m)},J_{\Omega(m)}) \}^{1/2}, 
\label{ineq:DTF}
\end{eqnarray} 
where we have used the Schwarz inequality. Taking the thermodynamic limit 
of the system, we have 
${\displaystyle \lim_{B\downarrow 0}\lim_{\Lambda\uparrow\L}
D_{\Lambda,B} (A,A) = D(A)}$ and hence from~(\ref{ineq:DTF}) we obtain 
\begin{equation}
D (J_f) 
 \le {1\over {R_0}^2} \sum_{n=0}^{R_0-1} \sum_{m=0}^{R_0-1} 
      \{ D (J_{\Omega(n)}) D (J_{\Omega(m)}) \}^{1/2}. 
\label{ineq:DTF2}
\end{equation} 
The function $D(A)$ can be written as $D(A)=2\int^\infty_0 d\lambda 
\{\omega (AA(i\lambda)) - \omega^2(A) \}$ for an arbitrary self-adjoint 
operator $A$, where $A(i\lambda)$ denotes the time-evolved operator of $A$. 
Hence, in the large $R$ limit, $D(J_{\Omega(n)})$ relates to 
the uniform susceptibility in the form 
\begin{equation}
\chi_J = \lim_{R_0 \uparrow\infty} {1\over |\Omega(n)|} D(J_{\Omega(n)}).
\end{equation}
(See also Appendix.) 
For sufficiently large $R$, using $|\Omega(n)|=(2R_0 + 2n +1)^d$ and 
$R_0=R+r$, we have 
\begin{equation}
D(J_{\Omega(n)}) = \{(2R_0+2n+1)^d + o(R^d) \} \chi_J \le 4^d (R+r)^d \chi_J
\end{equation}
and hence, from~(\ref{ineq:DTF2}), we obtain  
\begin{equation}\label{ineq:DTF3}
D(J_f) \le 4^d (R+r)^d \chi_J. 
\end{equation}

Next, we discuss other parts in the right-hand side of~(\ref{ineq:1}). 
Since calculations of 
$\omega ( \{\Delta A_{\Omega_{\rm S}}^*,\Delta A_{\Omega_{\rm S}} \} )$ 
have been published 
in ref.~\cite{Martin}, we adopt the results and do not 
repeat the calculations here. Thereby we have an upper bound 
\begin{equation}\label{ineq:ACR}
\omega ( \{ \Delta A_{\Omega_{\rm S}}^*, \Delta A_{\Omega_{\rm S}} \} ) 
 \le \left\{ 
\begin{array}{ll}
O(R^{-\delta})  & \mbox{\hspace{1cm}} (0<\delta<d) \\%[0.5cm]
O(R^{-d}\ln R)  & \mbox{\hspace{1cm}} (\delta=d)   \\%[0.5cm]
O(R^{-d})       & \mbox{\hspace{1cm}} (\delta>d), 
\end{array} \right.
\end{equation}
where we have used the clustering property~(\ref{cond:cluster}).
Calculations of $\omega ( [[J_f,H],J_f] )$ are also given 
in ref.~\cite{Martin}. Though the definition of the smooth function 
$f(x)$ is different from ours, the derivations and results of 
ref.~\cite{Martin} still hold only by changing the spherical supports 
to the hyper-cubic ones. Thus we have 
\begin{equation}\label{ineq:DoubleC}
\omega ( [[J_f,H],J_f] ) \le M \Vert J_0 \Vert^2 R^{d-2} 
\sum_x |x|^2 \psi (x), 
\end{equation}
where $M$ is a positive finite constant. 
If we use $J_\Omega$ instead of $J_f$ in~(\ref{ineq:DoubleC}), 
$\omega ([[J_\Omega,H],J_\Omega])$ can be bounded by the form $R^{d-1}$. 
Thus in (\ref{ineq:DoubleC}) the double commutator is 
better estimated due to the smooth action. 
Inserting~(\ref{ineq:DTF3})--(\ref{ineq:DoubleC}) 
into~(\ref{ineq:1}), we obtain~(\ref{Lemma}).~\qed
%\vspace{12mm}
\vspace{6mm}

{\it Proof of Theorem~1.}
Setting $d=1$ in Lemma, taking the $R\rightarrow\infty$ limit, and 
using (\ref{eq:deriv}) and (\ref{eq:translation}), 
one obtains (\ref{eq:Theorem1}) for any $\delta >0$, 
if $\chi_J <\infty$.~\qed
%\vspace{12mm}
\vspace{6mm}

{\it Proof of Theorem~2.}
Consider the case that all hypotheses of this theorem are satisfied 
and furthermore assume that the truncated two-point 
correlation function of $A$ decays faster than $1/|x|^{d-1}$, i.e., 
\begin{equation}\label{eq:assumption}
|\omega (A^* \tau_x (A)) - \omega (A^*)\omega (\tau_x(A))| \le 
o \left( {1 \over |x|^{d-1}} \right). 
\end{equation}
Here $o(|x|^{-d+1})$ denotes a number that is lower order than $|x|^{-d+1}$. 
Using (\ref{eq:assumption}) instead of the clustering 
property (\ref{cond:cluster}), one can obtain 
\begin{equation}
\omega(\{ \Delta A^*_{\Omega_{\rm S}}, \Delta A_{\Omega_{\rm S}} \})
\le o(R^{-d+1})
\end{equation}
instead of (\ref{ineq:ACR}). Thus, slightly modifying the proof of Lemma, 
we obtain 
\begin{equation}\label{contradict}
\left| {1\over |\Omega_{\rm s}|} \sum_{x\in\Omega_{\rm s}} 
\omega ([J_\Omega,\tau_x(A)]) \right|^2
\le o(R^0), 
\end{equation}
where $o(R^0)$ denotes a number that vanishes in the $R\rightarrow \infty$ 
limit. 
(Remember that we are in the condition $\chi<\infty$.)
In the $R\rightarrow\infty$ limit, (\ref{contradict}) shows 
$\omega ([J_\Lambda,A])=0$. This clearly contradicts with the 
condition $\omega([J_\Lambda,A]) \ne 0$ and hence, by contradiction, 
we arrive at (\ref{eq:Theorem2}).~\qed

\section*{Acknowledgments}
The author would like to thank Professor~K.~Kubo for critically reading 
this manuscript and useful comments, and 
Professor~S.~Takada and Dr.~T.~Koma for stimulating discussions. 
He also acknowledges the financial support by the Japan Society 
for the Promotion of Science (JSPS). 
%\newpage

\renewcommand{\thesection}{\Alph{section}}
\setcounter{section}{0}
\section{Definitions of the uniform susceptibility}
We give a comment on the definition of the uniform susceptibility 
(\ref{def:suscep2}). 
In the literature, the susceptibility is usually defined by 
\begin{equation}\label{def:suscep1}
X_J\equiv\lim_{B\downarrow 0} \lim_{\Lambda\uparrow \L}
{1\over |\Lambda|} D_{\Lambda,B}(J_\Lambda), 
\end{equation}
where 
\begin{equation}
D_{\Lambda,B} (A) = \lim_{\beta\uparrow\infty}
\int^\beta_0 d\lambda 
\{ \omega_{\Lambda,B}^\beta (A^* A(i\lambda)) 
  - \omega_{\Lambda,B}^\beta (A^*) \omega_{\Lambda,B}^\beta (A) \}
\end{equation}
with 
\begin{equation}
\omega_{\Lambda,B}^\beta (\cdots) 
= { {\rm Tr} [\cdots \exp\{-\beta(H_\Lambda - B O_\Lambda) \}] \over 
    {\rm Tr} [\exp\{-\beta(H_\Lambda - B O_\Lambda) \}] }.
\end{equation}
For an arbitrary self-adjoint operator $A\in\A_\Lambda$, $D_{\Lambda,B} (A)$
can be written as 
\begin{equation}
D_{\Lambda,B} (A) = 
2 \int^\infty_0 d\lambda 
\{ \omega^{\beta=\infty}_{\Lambda,B} (A A(i\lambda)) 
  - \omega^{\beta=\infty}_{\Lambda,B} (A) 
    \omega^{\beta=\infty}_{\Lambda,B} (A) \}. 
\end{equation}
In (\ref{def:suscep1}), the limits are taken so that the state 
$\omega^\beta_{\Lambda,B}(\cdots)$ converges. 
Here we assume that the limits of the quantity in (\ref{def:suscep1}) 
exist and that $X_J$ is well-defined. 
Our definition of the uniform susceptibility is however different from 
(\ref{def:suscep1}). In this paper, we have defined the uniform susceptibility
as follows
\begin{eqnarray}
\chi_J &\equiv& \lim_{\Omega\uparrow\L} \lim_{B\downarrow 0} 
\lim_{\Lambda\uparrow \L} 
{1\over |\Omega|} D_{\Lambda,B}(J_\Omega) \nonumber\\
&=& \lim_{\Omega\uparrow\L} {2 \over |\Omega|}
\int^\infty_0 d\lambda \{ \omega (J_\Omega J_\Omega(i\lambda)) 
                          - \omega (J_\Omega) \omega (J_\Omega) \} 
\end{eqnarray}
taking suitable subsequences of $\Lambda$ and $B$, where $\Omega$ is set 
as the hyper-cubic subsets $\{ x\in\L$~:~$|x_i|\le R_0$~for~$i=1,\dots,d \}$ 
and 
\begin{equation}
\omega (\cdots) 
 = \lim_{B\downarrow 0} \lim_{\Lambda\uparrow\L} 
   \lim_{\beta\uparrow\infty} \omega_{\Lambda,B}^\beta (\cdots). 
\end{equation}
In this Appendix, we shall show that these two definitions are equivalent and 
hence $\chi_J$ converges to $X_J$.

Consider a finite subset $\Lambda(\supset\Omega)$ and a function $g(x)$ 
defined by 
\begin{equation}
g(x) = \left\{ 
\begin{array}{ll}
1  & \mbox{\hspace{1cm}} (x\in\Omega) \\%[0.5cm]
0  & \mbox{\hspace{1cm}} (x\notin\Omega), 
\end{array} \right.
\end{equation}
then we have $J_\Omega = \sum_{x\in\Lambda} g(x) J_x$ and 
\begin{eqnarray}
{1\over|\Omega|} D_{\Lambda,B} (J_\Omega)
&=& {1\over|\Omega|} {1\over|\Lambda|} D_{\Lambda,B} (\sum_k g_{-k} J_k) 
\nonumber\\
&=& {1\over|\Omega|} {1\over|\Lambda|} \sum_k |g_k|^2 D_{\Lambda,B} (J_k), 
\label{eq:relation}
\end{eqnarray}
where 
$J_k = |\Lambda|^{-1/2} \sum_{x\in\Lambda} J_x \exp(ikx)$ and 
$g_k = \sum_{x\in\Lambda} g(x) \exp(ikx)$. 
In the thermodynamic limit, (\ref{eq:relation}) can be written as 
\begin{equation}\label{eq:relation2}
\lim_{B\downarrow 0} \lim_{\Lambda \uparrow \L} 
{1\over|\Omega|} D_{\Lambda,B}(J_\Omega) 
= {1\over|\Omega|} \int_{|k_i|\le \pi} {d^d k \over (2\pi)^d} |g_k|^2 X_J (k),
\end{equation}
where 
\begin{equation}
X_J (k) = \lim_{B\downarrow 0} \lim_{\Lambda\uparrow \L} D_{\Lambda,B} (J_k). 
\end{equation}
The function $|\Omega|^{-1} |g_k|^2$ has the following two properties; 
\begin{equation}
\int_{|k_i|\le \pi} {d^d k \over (2\pi)^d} {1\over |\Omega|} |g_k|^2 = 1
\end{equation}
and 
\begin{eqnarray}
\lim_{\Omega\uparrow\L} {1\over |\Omega|} |g_k|^2 
&=& \lim_{R_0 \uparrow \infty} {1\over (2R_0+1)^d} 
\left\{ \prod^d_{i=1} {\sin k_i ( R_0+1/2 ) \over \sin k_i/2} \right\}^2 
\nonumber\\
&=& 0
\end{eqnarray}
for any $k$ satisfying $k\ne 0$ and $|k_i|\le\pi$. 
Hence it converges to the Dirac's delta function, 
\begin{equation}\label{eq:delta}
\lim_{\Omega\uparrow\L} {1\over|\Omega|} |g_k|^2 = (2\pi)^d \delta (k)
\end{equation}
for $|k_i|\le\pi$. Inserting (\ref{eq:delta}) into (\ref{eq:relation2}) and 
using $X_J (k=0) = X_J$, we thus find that $\chi_J = X_J$.

\end{document}